\documentclass[aps,prl,preprint,superscriptaddress,showpacs]{revtex4}

\usepackage{graphicx}

\begin{document}

\title{Quantum Magnetic Deflagration in
Mn$_{12}$ Acetate}

\author{A. Hern\'{a}ndez-M\'{i}nguez, J. M. Hernandez,  F. Maci\`{a}, A. Garc\'{i}a-Santiago, J. Tejada}

\affiliation{Departament de F\'{i}sica Fonamental, Facultat de
F\"{i}sica, Universitat de Barcelona, Avda. Diagonal 647, Planta
4, Edifici nou, 08028 Barcelona, Spain}

\author{P. V. Santos}

\affiliation{Paul-Drude-Institut f\"{u}r Festk\"{o}rperelektronik,
Hausvogteiplatz 5-7, 10117 Berlin}

\date{\today}

\begin{abstract}

We report controlled ignition of magnetization reversal avalanches
by surface acoustic waves in a single crystal of Mn$_{12}$
acetate. Our data show that the speed of the avalanche exhibits
maxima on the magnetic field at the tunneling resonances of
Mn$_{12}$. Combined with the evidence of magnetic deflagration in
Mn$_{12}$ acetate (Suzuki \textit{et al}., cond-mat/0506569) this
suggests a novel physical phenomenon: deflagration assisted by
quantum tunneling.
\end{abstract}


\maketitle

Magnetic properties of Mn$_{12}$-acetate have been intensively
studied after the magnetic bi-stability of this molecular cluster
below 3.5 K was demonstrated \cite{Sessoli}. The bi-stability is
caused by a large spin of the cluster, $S = 10$, and by strong
uniaxial magnetic anisotropy that provides a 65 K energy barrier
between spin-up and spin-down states. At low temperature a
magnetized Mn$_{12}$ crystal exhibits two modes of magnetic
relaxation. The first mode is a slow one. It manifests itself in a
staircase hysteresis curve which is due to thermally assisted
quantum tunneling of the magnetization \cite{Friedman}. The second
relaxation mode, exhibited by sufficiently large crystals, is a
much more rapid magnetization reversal that typically lasts less
than 1 ms. It was initially studied by Paulsen and Park
\cite{Paulsen} and attributed to a thermal runaway or avalanche
(see also Ref. \onlinecite{avalanches}). In the avalanche, the
initial relaxation of the magnetization towards the direction of
the field results in the release of heat that further accelerates
the magnetic relaxation. Recent local magnetic measurements of
Mn$_{12}$ crystals \cite{Suzuki} have demonstrated that during an
avalanche the magnetization reversal occurs inside a narrow
interface that propagates through a crystal at a constant speed of
a few meters per second. It has been argued that this process is
analogous to the propagation of a flame front (deflagration)
through a flammable chemical substance. The conventional theory of
deflagration, in the first approximation, yields the following
expression for the velocity of the flame front
\cite{LL,Combustion,Suzuki}:
\begin{equation}
v =
\sqrt{\frac{\kappa}{\tau_0}}\exp{\left(-\frac{U}{2k_BT_f}\right)}\,.
\end{equation}
Here $U$, $\tau_0$, and $T_f$ are the energy barrier, the attempt
frequency, and the temperature of the ``flame" in the expression
$\tau = \tau_0 \exp{\left({U}/{k_BT_f}\right)}$ for the ``chemical
reaction" time, and $\kappa$ is thermal diffusivity. In the case
of Mn$_{12}$, $\kappa \sim$ $10^{-5}\,$ m$^2$/s, $\tau_0 \sim
10^{-7}\,$s, and the field dependence of the energy barrier,
$U(H)$, is well known.

\begin{figure}
\includegraphics[width=4in,angle=-90]{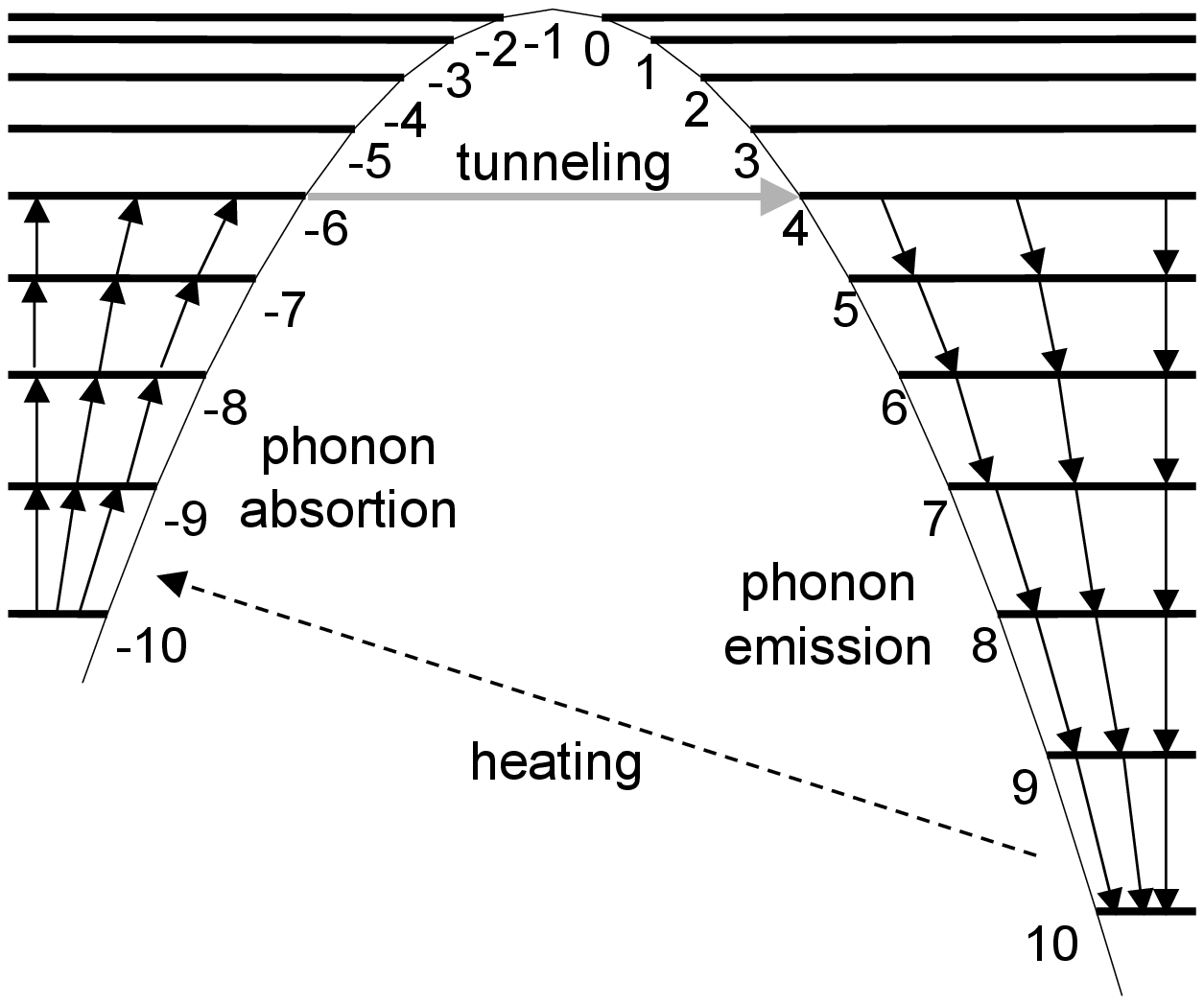}
\caption{Spin levels and magnetic avalanche in Mn$_{12}$ acetate. At
resonant fields, quantum tunneling effectively cuts out the top of
the barrier, increasing the rate of relaxation between the left and
right wells. Spin in the right well relax down the well releasing
phonons that produce heat, which is absorbed by spins in the left
well producing the avalanche.}
\end{figure}

In a flammable chemical substance the potential barrier is a
constant determined by the nature of the chemical reaction that
transforms a metastable chemical into a stable chemical
(\textit{e.g.} a mixture of hydrogen and oxygen transforms into
water). On the contrary, in molecular magnets the energy barrier,
as well as the released energy, can be controlled by the magnetic
field. At certain values of the magnetic field the spin levels on
the two sides of the energy barrier come to resonance and
thermally assisted quantum spin tunneling under the barrier takes
place, see Fig. 1. Therefore the effect of the tunneling is
roughly equivalent to cutting out the top of the barrier. This
well understood effect is responsible for the staircase hysteresis
curve in Mn$_{12}$ and other molecular magnets. Due to this
effect, one also should expect that the velocity of the avalanche,
given by Eq. (1), increases at the resonant values of the magnetic
field.

In the experiment of Suzuki \textit{et al}. \cite{Suzuki}
avalanches were ignited in a stochastic way on sweeping magnetic
field between $-5\,$T and $5\,$T. In such an experiment one cannot
control the ignition process. Consequently, the probability that
the avalanche occurs at a tunneling resonance is low, which may
explain why no oscillation of $v$ on $H$ due to resonant spin
tunnelling has been observed. In this Letter we report a novel
method of controlled ignition of avalanches in Mn$_{12}$ acetate
at constant magnetic field by means of surface acoustic waves
(SAW). We demonstrate that the velocity of the deflagration front,
and the ignition rate, oscillate on the magnetic field in
accordance with the expectation that quantum spin tunneling lowers
the barrier for the deflagration. Thus, in effect, Mn$_{12}$
acetate exhibits a phenomenon which has not been seen in any other
substance: slow burning (deflagration) assisted by quantum
tunneling.

\begin{figure}
\includegraphics[width=4in]{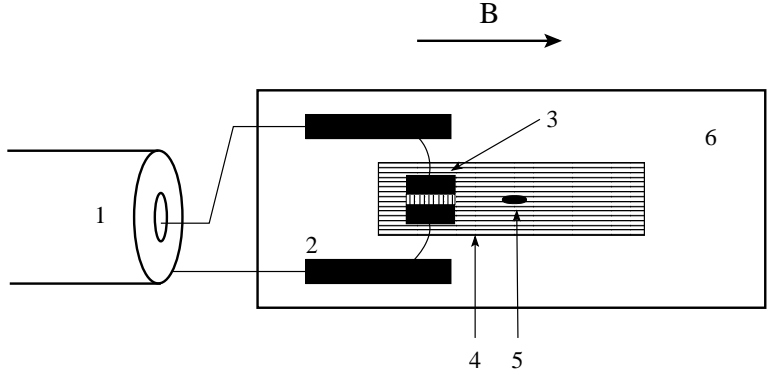}
\caption{Experimental set-up. The microwave signal is taken by a
coaxial cable (1) through a pair of conducting stripes (2) to the
transducers (3), which generate the surface acoustic waves on the
LiNbO$_3$ substrate (4). The Mn$_{12}$ crystal (5) is glued directly
on the substrate, with the easy anisotropy axis parallel to the
magnetic field direction. The whole assembly is mounted on a plastic
stand (6).}
\end{figure}

The acoustomagnetic experiments were performed by using hybrid
piezoelectric interdigital transducers (IDT) deposited on the 128 YX
cut of a LiNbO$_3$ substrate, Fig. 2. The transducers
\cite{Yamanouchi} generate multiple harmonics of the fundamental
frequency $111\,$MHz up to a maximum frequency of approximately
$1.5\,$GHz. A single crystal of Mn$_{12}$ acetate was glued directly
onto the IDT, using commercial silicon grease.

Experiments have been carried out inside a commercial rf-SQUID
Quantum Design Magnetometer. The microwaves for the SAW generation
were transported to the transducers with the help of coaxial
cables which introduce attenuation smaller than $10\,$dB. We
measured simultaneously the reflection coefficient $S_{11}$ and
the magnetization of the Mn$_{12}$ crystals as a function of
temperature ($2-300\,$K) and magnetic field up to $5\,$T. During
the experiments, the temperature of the IDT attached to the sample
and the temperature of the helium gas that provided heat exchange
were independently monitored. IDTs with and without Mn$_{12}$
crystals showed the same acoustic resonance frequencies that
weakly depend on temperature and magnetic field. The magnetization
dynamics was generated by exciting the IDTs with microwave pulses
from a commercial Agilent signal generator. This generator permits
selection of the shape, duration, and energy of the pulse in the
frequency range $250\,$kHz$- 4\,$GHz. Most of the experiments were
performed using rectangular microwave pulses of duration
$10\,\mu$s - $10\,$ms. The measurement of the reflection
coefficient $S_{11}$ was performed by using an Agilent
network analyzer. Fast magnetization measurements (time resolution
of $1\,\mu$s) were carried out at a constant temperature and
constant magnetic field by continuously reading the voltage
variation detected by the rf SQUID.

The magnetic properties of the Mn$_{12}$ single crystals glued
onto the surface of the transducer were found to be similar to
those previously published
\cite{Friedman,CT,Barbara,GS,Jonathan,Barco}. Above the magnetic
blocking temperature, $T_B$, the material exhibits
superparamagnetic behavior with the magnetization following a
$1/T$ law and a Brillouin dependence as a function of the ratio
$H/T$. Below $T_B$, the magnetization as a function of the
magnetic field shows resonant spin tunneling transitions at
magnetic field values multiple of $0.45\,$T. In order to estimate
the energy deposited by the SAW into the Mn$_{12}$ crystal, we
used microwave pulses of different duration in the
superparamagnetic regime, $T > T_B$, where $M(H)$ is a known
function determined by the thermal equilibrium among the spin
states.  The temperature variation of the Mn$_{12}$ crystal,
induced by the SAW, ranged from $0.1\,$K  to $1\,$K.

To study magnetic avalanches, we first saturated the sample at
$-2\,$T and $2.1\,$K. The field was then varied at a constant rate
of $300\,$Oe/s until a desired value of $H$ was reached. Maintaining
this field value, we delivered to the sample the SAW pulses of
increasing duration until the avalanche was triggered. This way, for
each value of the magnetic field, a threshold duration of the SAW
pulse needed to ignite the avalanche (that is, the threshold energy
delivered to the sample) has been established. The required duration
of the pulse was found to decrease with the magnetic field in
accordance with the expectation that the ``flammability" of the
Mn$_{12}$ crystal increases on increasing $H$. These measurements
have established a new method of igniting magnetization avalanches
in molecular magnets. The advantage of using SAW is a total control
over the field at which the avalanche takes place.

\begin{figure}
\includegraphics[width=4in,angle=-90]{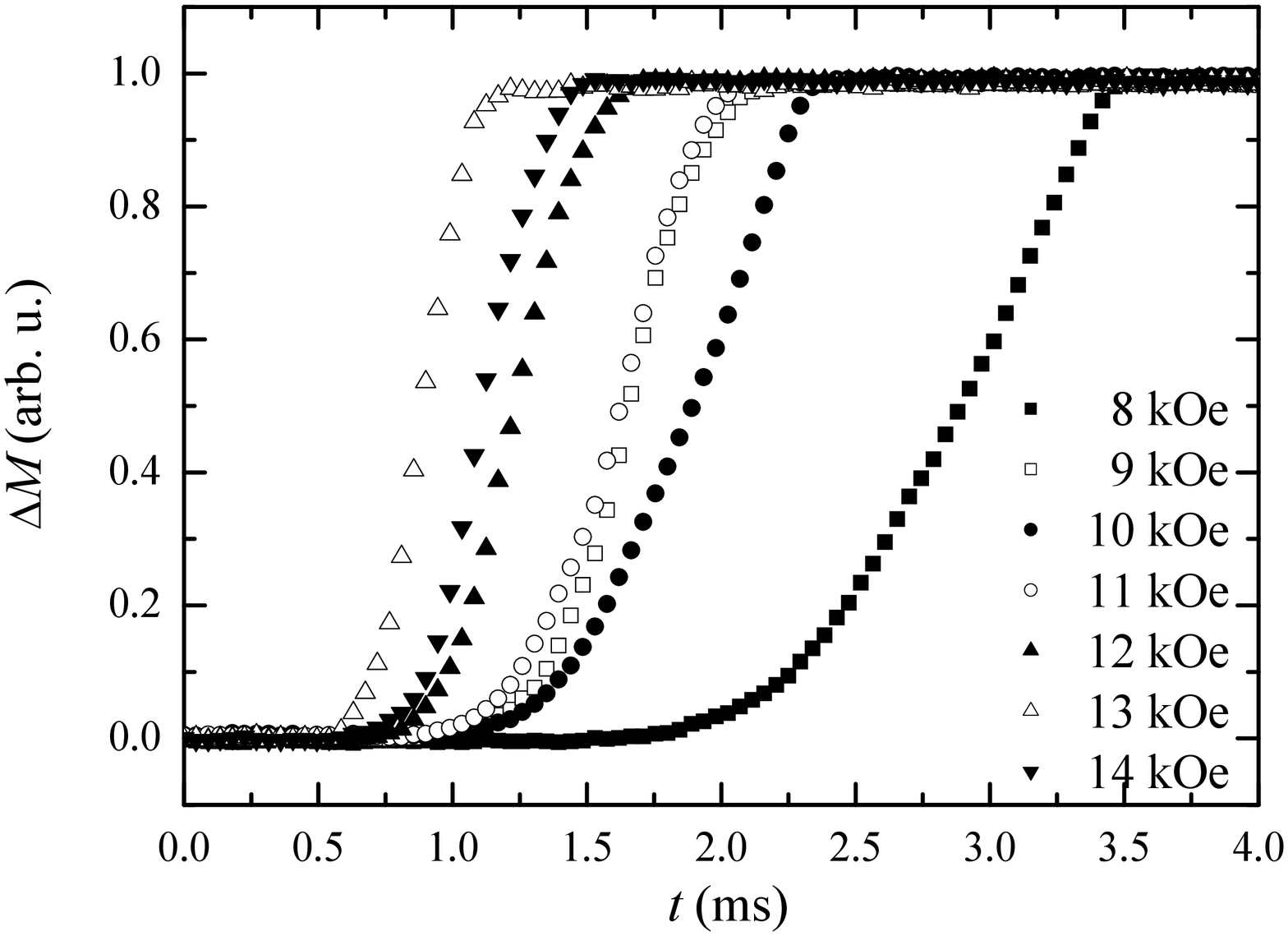}
\caption{Time dependence of the magnetization change during the
avalanche at different magnetic fields.}
\end{figure}

Fig. 3 shows the time dependence of the magnetization change
during avalanches triggered at different magnetic fields. The
speed, $v$, of the developed deflagration front can be obtained
from the data as $v = l/t$, where $t$ is the duration of the
avalanche and $l$ is the length of the sample. The dependence of
$v$ on $H$, extracted from the data, is shown in Fig. 4. The
dotted line represents the fit of the data by Eq. (1) with the
effective energy barrier cut by quantum tunneling near resonance
fields. The field dependence of the effective barrier was taken
from earlier experiments on magnetic relaxation \cite{EPL,SH}. The
temperature of the flame, $T_f$, was obtained using measured
specific heat of Mn$_{12}$ acetate \cite{Novak,Fominaya}. $T_f$ is
related to the specific heat, $C(T)$, through the relation
\begin{equation}
H \Delta M = \int_{T_i}^{T_f} dT C(T)\,,
\end{equation}
where $H$ is the field at which the avalanche occurs, $\Delta M$ is
the total change of magnetization, and $T_i$ is the initial
temperature. $T_f$ was found to be in the range between $6\,$K and
$9\,$K, depending on the magnetic field. The only fitting parameter
was $\sqrt{\kappa \tau_0}$ for which the value of $2 \times 10^{-6}\,$m
was obtained. This agrees with $\kappa \sim 10^{-5}\,$m$^2$/s and
$\tau_0 \sim 10^{-7}\,$s, known from independent measurements.

\begin{figure}
\includegraphics[width=4in,angle=-90]{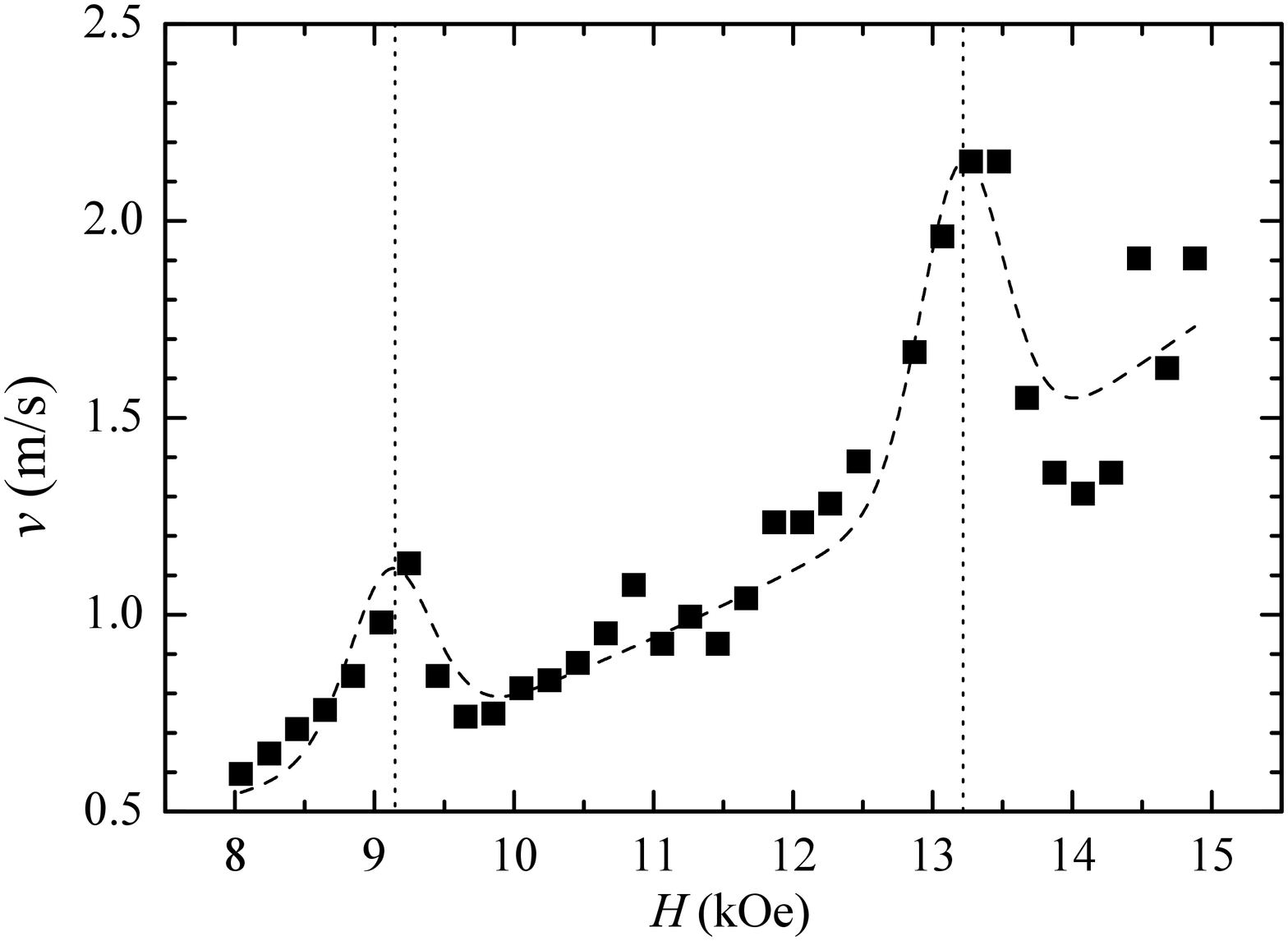}
\caption{Magnetic deflagration speed as function of the magnetic
field. The dashed line represents the fit by Eq. (1). Vertical lines
show the positions of the tunneling resonances.}
\end{figure}

Vertical lines in Fig. 4 represent positions of the tunneling
resonances for Mn$_{12}$ acetate. It is clear from the figure that
the maxima in $v(H)$ correspond to the tunneling resonances. This
observation is a clear evidence of the quantum aspect of the
deflagration. To be precise, the magnetic deflagration corresponding
to the avalanche is, of course, a thermal phenomenon, driven by
thermal conductivity. However, the speed of the deflagration is also
determined by the speed of ``chemical reaction", which, in our case,
is the rate of the transition between the two wells shown in Fig. 1.
This rate is determined by both thermal activation to the excited
spin levels and quantum tunneling between the levels. It is actually
the thermally assisted quantum tunneling that accelerates
deflagration at tunneling resonances. To our knowledge, slow burning
(deflagration) assisted by quantum tunneling has never been observed
in any chemical substance. In a crystal of molecular nanomagnet it
becomes observable due to the possibility to control the speed of
the deflagration by the magnetic field.

In conclusion, a new method of igniting magnetization reversal in
molecular magnets by surface acoustic waves has been developed. We
have observed a fundamentally new phenomenon: magnetic deflagration
assisted by quantum tunneling. We have demonstrated that quantum
tunneling can be seen not only in slow relaxation of molecular
magnets (through staircase hysteresis loop) but also in the fast
relaxation, that is avalanche, which is a process equivalent to slow
burning. This observation opens a new way for the study of quantum
phenomena in molecular magnets, as well as for the experimental
study of the complex deflagration physics.

\begin{acknowledgments}
We thank E.M. Chudnovsky and M.P. Sarachick for very helpfull discussions.
A.H.-M. and thanks the Spanish Ministerio de Educacion y Ciencia for a research grant.
J.M. H. thanks the Ministerio de Educacnion y Ciencia and the Univ. of Barcelona for a Ramon y Cajal research contract. A. G.-S. thanks the Univ. of Barcelona for an assistant professor contract. F.M. and J.T. thank SAMCA Enterprise for financial support.
 
\end{acknowledgments}

\end{document}